\documentclass[10pt,aps,twocolumn,superscriptaddress,preprintnumbers,nofootinbib]{revtex4-1} 
\usepackage{amsmath,amssymb,bbold,graphicx,natbib,bm}
\usepackage{enumerate}
\usepackage{hyperref}
\usepackage{color}

\usepackage{empheq}
\usepackage{slashed}

\newcommand{\beq}{\begin {equation}}  
\newcommand{\eeq}{\end   {equation}} 
\newcommand{\bea}{\begin {eqnarray}} 
\newcommand{\eea}{\end   {eqnarray}}  
\newcommand{\baa}{\begin {array}   } 
\newcommand{\eaa}{\end   {array}   }     
\newcommand{\bit}{\begin {itemize} }
\newcommand{\eit}{\end   {itemize} }
\newcommand{\be }{\begin {equation}} 
\newcommand{\ee }{\end   {equation}}
\newcommand{\nn }{\nonumber        }

\newcommand{\bv}{\ensuremath{\mathbf{v}}}  
\newcommand{\vmin}{\ensuremath{v_\textrm{min}}}

\def\mag{\tilde{\mu}}
\def\W{\tilde{W}}

\def\lsim{\mathrel{\rlap{\lower4pt\hbox{$\sim$}}
   \raise1pt\hbox{$<$}}}                
\def\gsim{\mathrel{\rlap{\lower4pt\hbox{$\sim$}}
   \raise1pt\hbox{$>$}}}                

\begin{document}


\preprint{UTTG-20-14, TCC-022-14}

\title{Lepton-flavored asymmetric dark matter and interference in direct detection}

\author{Ali Hamze}
\author{Can Kilic}
\author{Jason Koeller}
\author{Cynthia Trendafilova}
\author{Jiang-Hao Yu}
\affiliation{Theory Group, Department of Physics and Texas Cosmology Center,
\\The University of Texas at Austin,  Austin, TX 78712 U.S.A.}


\begin{abstract}
In flavored dark matter models, dark matter can scatter off of nuclei through Higgs and photon exchange, both of which can arise from renormalizable interactions and individually lead to strong constraints from direct detection. While these two interaction channels can destructively interfere in the scattering amplitude, for a thermal relic with equal abundances for the dark matter particle and its antiparticle, this produces no effect on the total event rate. Focusing on lepton-flavored dark matter, we show that it is quite natural for dark matter to have become asymmetric during high-scale leptogenesis, and that in this case the direct detection bounds can be significantly weakened due to interference. We quantify this by mapping out and comparing the regions of parameter space that are excluded by direct detection for the symmetric and asymmetric cases of lepton-flavored dark matter. In particular, we show that the entire parameter region except for a narrow Higgs resonance window is ruled out in the symmetric case for fermion dark matter when the coupling to the Higgs dominates over the coupling to leptons, while large portions of parameter space are still allowed for the asymmetric case. The same is also true for a dark matter mass above 8~GeV for scalar dark matter when the coupling to leptons dominates over the coupling to the Higgs.
\end{abstract}

\maketitle


\section{Introduction}
\label{sec:intro}

The steady improvement in the sensitivity of direct detection searches is putting severe constraints on the parameter space of dark matter (DM) models belonging to the weakly interacting massive particle (WIMP) paradigm. These bounds can be relaxed in certain classes of models, including Majorana fermion DM where only spin-dependent scattering contributes, inelastic DM~\cite{TuckerSmith:2001hy, TuckerSmith:2004jv, Chang:2008gd} where the observed event rate is severely reduced due to the energy cost of upscattering, or isospin violating DM\cite{Feng:2011vu, Chang:2010yk} where destructive interference can occur between the scattering of DM off of protons and neutrons, among others. The idea of destructive interference in the scattering amplitude has been used in several dark matter models in the past~\cite{Fox:2011qd, Gao:2011ka, DelNobile:2011yb, DelNobile:2011je, Frandsen:2011cg, Cirigliano:2013zta, Okada:2013cba, Belanger:2013tla, Hamaguchi:2014pja, Chen:2014cbt}. A simple class of models that can give rise to interference is when the DM particle interacts with nuclei via multiple mediators. A nontrivial check in such models is whether the parameters of the model need to be fine-tuned, or in other words, whether scattering amplitudes for the exchange of the mediators are naturally of the same size for generic values of the couplings in the model.
	
In this paper we argue that flavored dark matter (FDM) models~\cite{Kile:2011mn, Batell:2011tc, Kamenik:2011nb,Agrawal:2011ze,  Kumar:2013hfa, Chang:2013oia, Bai:2013iqa,Batell:2013zwa,Agrawal:2014aoa,Agrawal:2014una,Agrawal:2014ufa} can give rise to interference in the scattering amplitude quite naturally. These models admit renormalizable couplings between the DM and SM fields that lead to both tree-level Higgs exchange as well as loop-level photon exchange channels for direct detection, with comparable sizes.
	
Unfortunately, interference between spin-0 (Higgs) and spin-1 (photon) mediated channels will not in general help to ease direct detection constraints for WIMPs, which have equal relic abundances for the DM particle $\chi$ and its antiparticle $\bar{\chi}$. The amplitude for a spin-0 exchange channel will have the same sign for $\chi$ and $\bar{\chi}$, while the amplitude for a spin-1 exchange channel will change sign\footnote{In FDM models, $\chi$ cannot be a self-conjugate field due flavor constraints. See Sec.~\ref{sec:model} for details.}. Therefore, any destructive interference that occurs for the scattering of $\chi$ off of nuclei will unavoidably lead to constructive interference in the scattering of $\bar{\chi}$, and the total scattering rate will be the same as in the absence of any interference.
	
On the other hand, for asymmetric DM~\cite{Nussinov:1985xr, Kaplan:1991ah, Barr:1990ca, Kaplan:2009ag, Falkowski:2011xh, Zurek:2013wia, Petraki:2013wwa}, the destructive interference can significantly weaken direct detection constraints. Interestingly, this too can occur readily in FDM models. In  this paper we  focus on the case of lepton-flavored DM, where we will show that it is very natural for a DM asymmetry to be generated during high-scale leptogenesis~\cite{Fukugita:1986hr} (for additional references see reviews on this subject, e.g.~\cite{ Buchmuller:2005eh, Chen:2007fv}). Using lepton-flavored asymmetric DM as our benchmark model, and contrasting with the same model but with a symmetric $\chi$-$\bar{\chi}$ abundance, we will quantify the impact of interference on the region of parameter space that is compatible with the null results of direct detection experiments. In particular, we will show that for the case of fermion dark matter that couples predominantly to the Higgs, the full parameter region in the symmetric case is ruled out due to direct detection except for a narrow Higgs resonance window, while the asymmetric case can be consistent with the bounds due to interference. This is also true for the case of scalar DM that couples predominantly to leptons when the DM mass is above 8~GeV.

The particle content of FDM models includes three copies of the DM particle $\chi$ as well as a mediator particle $\phi$ which makes renormalizable interactions between $\chi$ and the standard model (SM) fermions possible. Due to Lorentz invariance, one of $\chi$ and $\phi$ is necessarily a fermion while the other one is a boson. We will study both possibilities for completeness and highlight the similarities as well as the differences between them.

The outline of the paper is as follows: In Sec. \ref{sec:model} we will review the lepton-flavored DM model and describe its general features, before introducing a mechanism by which it can become asymmetric during high-scale leptogenesis. We will go over the direct detection prospects of lepton-flavored DM in Sec. \ref{sec:DD} and we will map out the excluded regions in the parameter space of the model for both the symmetric and asymmetric cases in section \ref{sec:results}. We will conclude in Sec. \ref{sec:conclusions} and comment on future directions. Detailed formulae related to the calculation of the relic density in the symmetric case and to the scattering amplitude for direct detection can be found in the appendices.


\section{The Model}
\label{sec:model}

The FDM setup has been described in detail in Ref.~\cite{Agrawal:2011ze} so we will only give a brief summary here. The DM is taken to be a singlet under the gauge symmetries of the standard model (SM) but it belongs to a multiplet that transforms nontrivially under the flavor symmetries of the SM, which we will denote by $\chi_{i}$. There is also a mediator particle $\phi$ which is a flavor singlet, but which carries SM hypercharge. Assuming that the $\phi$ mass is heavier than at least one of the $\chi$ masses, the lightest of the $\chi_{i}$ is rendered stable by a global $U(1)$ under which only the $\chi_{i}$ and $\phi$ are charged. We will refer to this $U(1)$ as $\chi$-number.

It was shown in Ref.~\cite{Agrawal:2011ze} that FDM is compatible with constraints arising from flavor observables in a minimal flavor violation (MFV)~\cite{D'Ambrosio:2002ex} setup, such that the SM Yukawa couplings are the only source of flavor violation. With this assumption, the minimal choice in terms of the number of degrees of freedom is for $\chi_{i}$ to be a flavor triplet.

Which SM flavor symmetry $\chi_{i}$ transforms under determines the SM fermions it can couple to at the renormalizable level. For the rest of this paper we will focus our attention on the specific case of lepton-flavored DM, where $\chi_{e,\mu,\tau}$ transform as a triplet under $SU(3)_{e_{R}}$. As in Ref.~\cite{Agrawal:2011ze}, we will work with a benchmark model where $\chi_{\tau}$ is the lightest state, but the main conclusions of this paper are insensitive to this choice. A renormalizable coupling to the SM fermions requires one of $\chi$ and $\phi$ to be a fermion, and the other to be a scalar. Note that in order to be a triplet under $SU(3)_{e_{R}}$, $\chi_{i}$ must be complex, so it is either a complex scalar or a Dirac fermion. If the DM is a scalar, the interaction term is
\bea
	{\mathcal L}_{\rm scalar} \supset \lambda_{ij} \chi_i \bar{\phi} e_{R,j}  + {\rm h.c.},
	\label{eq:FDMvertexS}
\eea
while for a fermionic DM it has the form
\bea
	{\mathcal L}_{\rm fermion} \supset \lambda_{ij} \bar{\chi}_i \phi e_{R,j}  +{\rm h.c.}.
	\label{eq:FDMvertexF}
\eea
As discussed in Ref.~\cite{Agrawal:2011ze}, within the MFV formalism the flavor structure of $\lambda_{ij}$ is
\bea
	\lambda_{ij} = (\alpha \mathbb{1} + \beta y^\dagger y)_{ij}.
\eea  
In order to reduce clutter, we will assume that $\alpha\gg\beta$, such that we can define $\lambda_{ij} \equiv \lambda_\phi \delta_{ij}$. It should be noted however that this is mainly a choice of convenience and that the main conclusions of this paper are not sensitive to this choice.

In the scalar DM case, the only other renormalizable interaction of the dark sector with the SM allowed by the symmetries of the model is a coupling to the Higgs doublet. Including this interaction, the scalar potential can be written as
\bea
	V_{\rm scalar} &=& \lambda_h (H^\dagger H - \frac12 v^2)^2 +  \mu_{\chi_i}^2 \chi_i^* \chi_i \nn\\
	&+&   \lambda_{\chi h} \chi_i^* \chi_i H^\dagger H + \lambda_s (\chi_i^* \chi_i)^2.
	\label{eq:scalarV}
\eea
This potential is bounded from below even for $\lambda_{\chi h}<0$, provided that
\bea
\lambda_h > 0,\qquad \lambda_s > 0,\qquad \lambda_h \lambda_s > \frac14 \lambda_{\chi h}^2.
\eea
Note that negative value $\lambda_{\chi h}$ does not present a problem as long as $\lambda_s$ is positive and large.
After electroweak symmetry breaking, the DM inherits a $\chi$-$\chi$-$h$ coupling. This will contribute to direct detection through tree-level Higgs exchange.

In order to study similar phenomenological features in the fermion DM case, we will also include a dimension-5 term in the Lagrangian
\bea
{\mathcal L}_{\rm fermion} \supset -\frac{\kappa}{\Lambda} \bar{\chi_i} \chi_i H^\dagger H.
\label{eq:FDMH}
\eea
To have consistency between the scalar and fermion DM cases, we will adopt a convention such that
\bea
\frac{\kappa}{\Lambda}\equiv\frac{\lambda_{\chi h}}{v},
\label{eq:dim5}
\eea
where $v$ is the electroweak scale, and
with the understanding that $\lambda_{\chi h}$ is small in the fermionic DM case. In other words, the dimension-5 term is assumed to have arisen by integrating out additional degrees of freedom at the scale $\Lambda$ (close to TeV scale), such as a heavy SM singlet scalar with couplings to $\bar{\chi}$-$\chi$, and to the SM Higgs. Note that the scalar potential in this case can also include a renormalizable $|\phi|^{2}|H|^{2}$ term, but the presence of this term will have no effect for the rest of the paper, and for this reason we will not dwell on it any further.

Let us now turn our attention to the generation of a $\chi$ asymmetry. We will demonstrate this explicitly in the fermion DM case; it is straightforward to implement the same mechanism in the scalar DM case as well. We assume that a primordial lepton asymmetry is generated via the decay of right-handed neutrinos at a high scale within a few orders of magnitude of the scale of grand unification. The right handed neutrinos $N_{R}$ couple to the SM leptons through 
\begin{eqnarray}
{\mathcal L}_{\rm lepton}&=&\frac{1}{2}(M_{N})_{ij}\overline{N}^{c}_{R,i} N_{R,j}\nonumber\\
&+&\left(y^{L}_{ij} \bar{L}_{i} H e_{R,j} + y^{N}_{ij} \bar{L}_{i}\tilde{H}N_{R,j} \right) + {\rm h.c.},
\label{eq:SMyukawa}
\end{eqnarray}
where $L_{i}$ are the $SU(2)$ doublet SM lepton fields, $\tilde{H}=\epsilon H^{*}$ and the first term is a Majorana mass for the right-handed neutrinos. The mechanism by which nonthermal decays of the right-handed neutrinos generate a nonzero lepton asymmetry, and later a nonzero baryon asymmetry through sphaleron processes, is well known (see~\cite{Buchmuller:2005eh, Chen:2007fv} and references therein). This mechanism relies on CP violating phases in the cross-terms between the tree-level and one-loop contributions to the amplitude for $N_{R}$ decay.

At first, it may seem that the interaction of Eq. \ref{eq:FDMvertexF} is sufficient to transfer any lepton asymmetry generated in the decays of $N_{R}$ to the $\chi_{i}$. However, $\chi$-number is still an exact symmetry at this point, which makes it impossible to generate a $\chi$ asymmetry from an asymmetry in a different species with no $\chi$-number. Therefore, the crucial ingredient for transferring the lepton asymmetry into the DM sector is breaking $\chi$-number (down to ${\mathbb Z}_{2}$ such that the stability of DM is not lost). For this purpose we add one more degree of freedom to the model, a {\it real} scalar field $S$, with the interaction
\begin{equation}
{\mathcal L}_{S}=y^{S}_{ij}\bar{\chi}_{i}S N_{R,j} + {\rm h.c.}.
\end{equation}
Since $S$ is real, this interaction breaks $\chi$-number, but there is still a ${\mathbb Z}_{2}$ under which $S$, $\phi$ and all three $\chi$ are odd. This interaction makes it possible for out-of-equilibrium decays of the right-handed neutrino to generate a $\chi$ asymmetry through interference between tree-level and one-loop contributions with CP violating phases, in the exact same way that the same decays also generate a lepton asymmetry. The couplings in ${\mathcal L}_{\rm fermion}$ which are assumed to be of order one will lead to efficient annihilation of the symmetric component of $\chi$. Note that there is no hierarchy problem associated with the scalar $S$, because it need not be light. The only requirement for this mechanism to work is for $S$ to not be heavier than the near-GUT scale right-handed neutrinos. Below the mass of $S$, $\chi$-number becomes an accidental symmetry. Both due to the presence of this symmetry, and due to the fact that both $\phi$ and $\chi$ are singlets under the weak $SU(2)$, the dark sector does not participate in the sphaleron processes which allow the original  asymmetry in the SM lepton sector to be transferred to the baryons.

Note that while the same mechanism   generates the lepton and $\chi$ asymmetries, the phases that determine the size of the generated asymmetry are different. In particular, the lepton asymmetry will depend on the physical combinations of phases in the matrices $y^{L}_{ij}$ and $y^{N}_{ij}$, whereas the $\chi$ asymmetry will depend on the phases in the matrices $\lambda_{ij}$ and $y^{S}_{ij}$. This means that if the phases that are relevant for the $\chi$ asymmetry are smaller than those that are relevant for the lepton asymmetry, the $\chi$ asymmetry will be smaller, and therefore $m_{\chi}$ must be chosen so that the $\chi$ energy density will be a factor of 5-6 larger than the baryon energy density. We will not assume any particular relation between the phases in the lepton and $\chi$ sectors, treating $m_{\chi}$ as a free parameter that is chosen such that $\chi$ has an energy density compatible with the DM density we observe in the universe today.

The collider phenomenology of asymmetric FDM is identical to the symmetric case, which was studied in Ref.~\cite{Agrawal:2011ze}, and we will not go into this in any further detail (See Sec. V for further comments). Any indirect detection signals for the symmetric case are of course nonexistent for the asymmetric case, so we will not have anything further to say about constraints from indirect detection either. In the rest of the paper we will concentrate on direct detection searches, where asymmetric FDM can have very different prospects compared to the symmetric case, due to the presence of interference, as we will study in detail in the next section.


\section{Direct Detection}
\label{sec:DD}

In this section we will calculate the cross section for $\chi$ to scatter off of an atomic nucleus, keeping interference terms. As mentioned in the introduction, when the DM is symmetric, the interference terms will cancel once the scattering of both $\chi$ and $\bar{\chi}$ are taken into account, but for asymmetric DM, they will be crucial. Based on the model of section~\ref{sec:model}, it is easy to see that scattering can happen at tree-level through Higgs exchange. At tree-level, the FDM interaction of Eqs.~(\ref{eq:FDMvertexS}) and (\ref{eq:FDMvertexF}) (for the scalar and fermion DM cases, respectively) does not contribute to the scattering, however as was studied in Ref.~\cite{Agrawal:2011ze}, it does give rise to vector exchange at loop order. The exchanged vector boson can be either the photon or the $Z$-boson, but of course the latter is strongly suppressed compared to the former due to the $Z$-mass. Therefore we will only consider the photon exchange for the rest of the paper.

\subsection{Scalar DM}

\begin{figure}[htp]
\begin{center}
\includegraphics[width=0.15\textwidth]{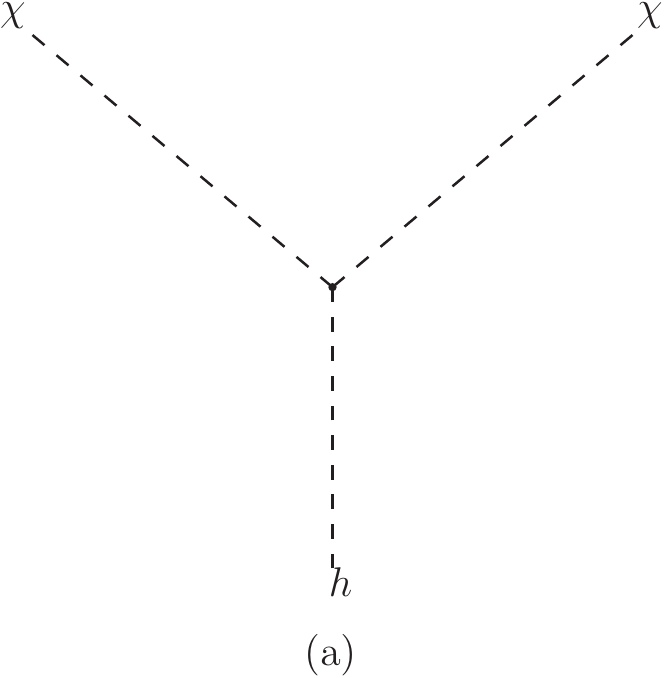} 
\includegraphics[width=0.15\textwidth]{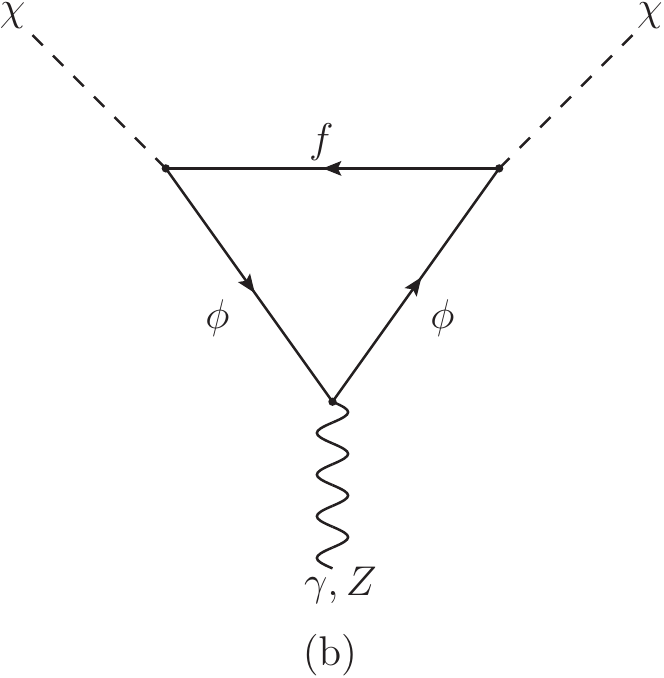} 
\includegraphics[width=0.15\textwidth]{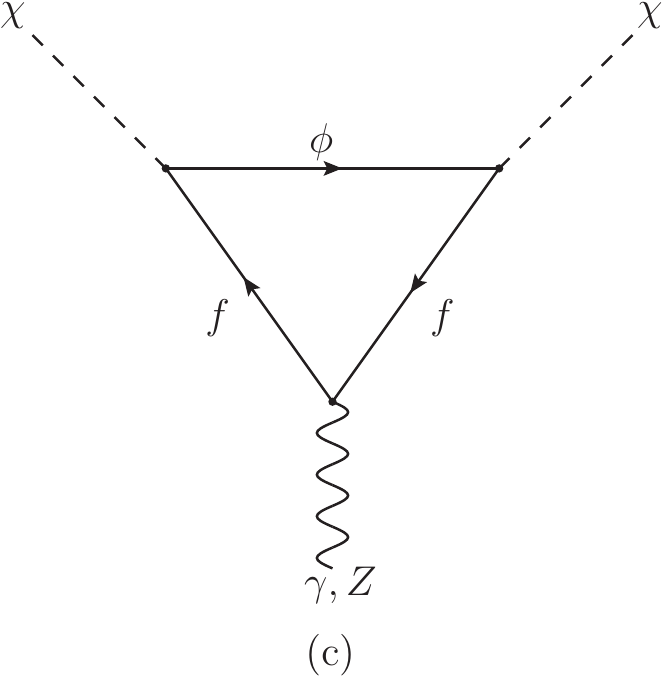} 
\caption{\small The Feynman diagrams that contribute to direct detection in the scalar DM case, namely tree level Higgs exchange (a) and $\gamma/Z$ exchange with the vector boson line attached to either the mediator $\phi$ (b) or to the SM fermions $f$ (c) running in the loop.}
\label{fig:vvvtri}
\end{center}
\end{figure}
	
After electroweak symmetry breaking, the interaction term in Eq.~(\ref{eq:scalarV}) contains the interaction
\begin{equation}
 \mathcal{L}_{h} \supset -v \lambda_{\chi h} \chi^* \chi h,
\end{equation}
which leads to the tree-level Higgs exchange. The loop-induced coupling of the DM to the photon is calculated in Appendix~\ref{app:loop} and in the zero external momentum limit it has the form
\bea
		b_\chi \partial^\mu \chi^* \partial^\nu \chi F_{\mu\nu},
\eea
where
\bea
	b_\chi \equiv - \frac{\lambda_\phi^2 e}{16 \pi^2 m_\phi^2} \left( 1 + \frac23 \ln\frac{m_\ell^2}{m_\phi^2}\right),
	\label{eq:sloopb}
\eea 
and $m_\ell$ is the mass of the tau lepton since we have assumed $\chi_{\tau}$ to be the DM.
The Feynman diagrams for these couplings are shown in figure~\ref{fig:vvvtri}.

Combining this with the Higgs and photon propagators, we can write the effective operators that give rise to the DM-nucleus scattering:
\begin{align}
				\mathcal{L}_{\rm eff} = c^q_\gamma \chi^\ast \overset{\text{\tiny{$\leftrightarrow$}}}{\partial}^\mu \chi \overline{q} \gamma_\mu q
				+ c^q_h \chi^\ast \chi \overline{q} q,			
\end{align}
where the coefficients are related to the couplings in the UV theory as
\begin{align}
	c^q_\gamma = e Q_q \frac{b_\chi}{2}, \quad
	c^q_h =  \frac{\lambda_{\chi h} m_q}{m_h^2}.
	\label{eq:SDMcq}
\end{align}

For the next step in calculating the scattering cross section, we convert from quark-level operators to effective nucleon-level operators and we take the nonrelativistic limit of the matrix elements, which gives ($N = p, n$)
\begin{align}
				\mathcal{L}_{\rm eff} = c^N_\gamma \chi^\ast \overset{\text{\tiny{$\leftrightarrow$}}}{\partial}^\mu \chi \overline{N} \gamma_\mu N
				+ c^N_h \chi^\ast \chi \overline{N} N.
				\label{eq:SDMLeffN}
\end{align}
The coefficients $c^N$ at the nucleon level can be written in terms of the coefficients $c^q$ at the quark level as
\begin{align}
				c^N_\gamma &= \frac{e  b_\chi}{2}\sum_q Q_q, \\
				c^N_h &= \sum_{q = u, d, s} c^q_h \frac{m_N}{m_q} f_{Tq}^{(N)} + \frac{2}{27} f_{TG}^{(N)}   \sum_{q = c, b, t} c^q_h \frac{m_N}{m_q},
\end{align}
where we use the numerical values of $ f_{Tq}^{(N)}$ and $f_{TG}^{(N)}$ given in Ref.~\cite{Belanger:2013oya}.
Combining with Eq.~(\ref{eq:SDMcq}) we arrive at
\begin{align}
				c^N_\gamma &=  \frac{e Q_N b_\chi}{2},\\
				c^N_h &=  \frac{\lambda_{\chi h} m_N}{m_h^2} \left(\frac{2}{9} + \frac{7}{9} \sum_{q = u,\,d,\,s} f_{Tq}^{(N)}\right).
\end{align}

The leading (spin-independent) contribution to the nucleon matrix elements of the operators of Eq.~(\ref{eq:SDMLeffN}) are
\begin{align}
				\langle \chi ,\, N \left| \chi^\ast \overset{\text{\tiny{$\leftrightarrow$}}}{\partial}^\mu \chi \overline{N} \gamma_\mu N \right| \chi ,\, N \rangle &= 4 m_\chi m_N ,\nonumber\\
				\langle \chi ,\, N \left| \chi^\ast \chi \overline{N} N \right| \chi ,\, N \rangle &= 2 m_N.
				\label{eq:NmatrixS}
\end{align}
Putting everything together, we define the dark matter-nucleon effective couplings
\begin{align}
	{\mathcal C}^N &=  4 m_\chi m_N c^N_\gamma + 2 m_N c^N_h,
	\label{eq:NeffS}
\end{align}
in terms of which the total scattering cross section is given by
\begin{equation}
				\sigma_T  = \frac{1}{16 \pi} \left( \frac{1}{m_\chi + m_p} \right)^2 \left[ Z {\mathcal C}^p + (A-Z) {\mathcal C}^n \right]^2.
				\label{eq:sfdmdd}
\end{equation}

\subsection{Fermion DM}

\begin{figure}[htp]
\begin{center}
\includegraphics[width=0.15\textwidth]{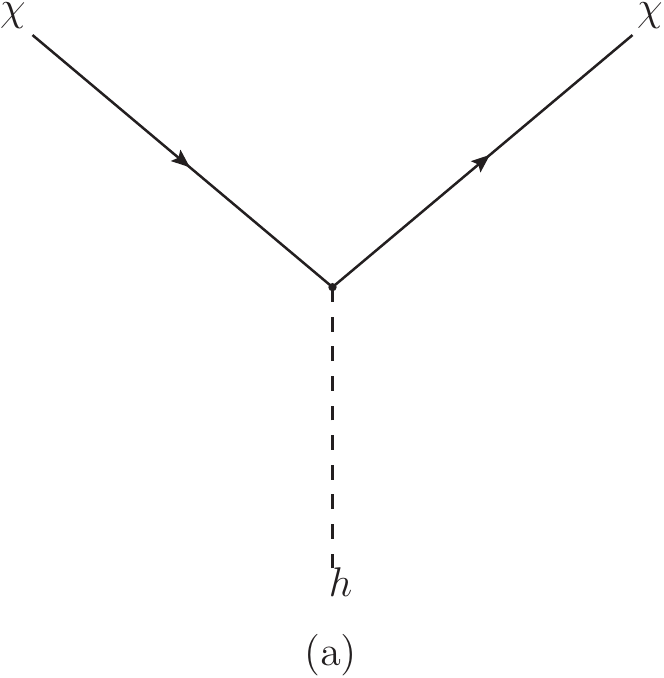} 
\includegraphics[width=0.15\textwidth]{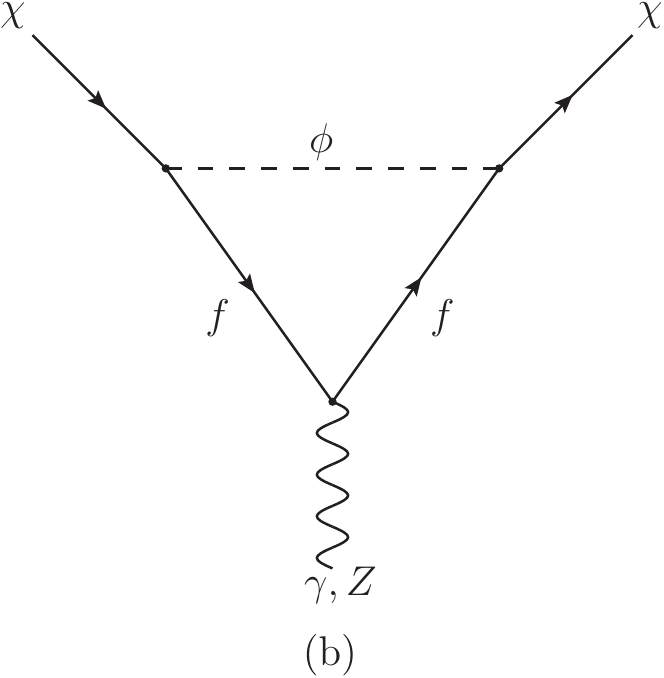} 
\includegraphics[width=0.15\textwidth]{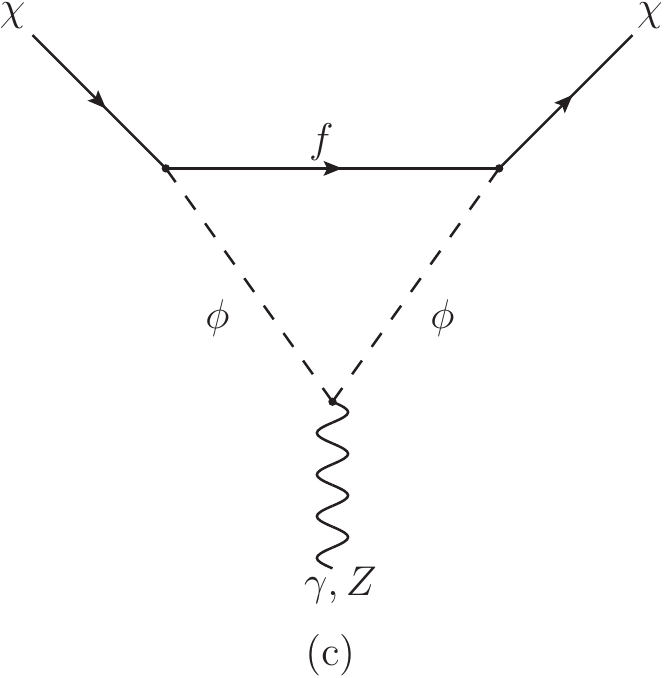} 
\caption{\small The Feynman diagrams that contribute to direct detection in the fermion DM case, namely tree level Higgs exchange (a) and $\gamma/Z$ exchange with the vector boson line attached to either the mediator $\phi$ (b) or to the SM fermions $f$ (c) running in the loop}
\label{fig:vvvtri2}
\end{center}
\end{figure}

The calculation of the scattering cross section for the fermion DM case proceeds through the same steps as in the scalar DM case. The tree-level Higgs exchange arises from the interaction of Eq.~(\ref{eq:FDMH}) after electroweak symmetry breaking
\begin{equation}
 \mathcal{L}_{h} \supset -   {\lambda_{\chi h}}  \bar{\chi} \chi h,
\end{equation}
while the loop induced coupling of the DM to the photon is given by
\bea
  {\mathcal L}_{\rm eff} &=& b_\chi  \bar{\chi} \gamma_\nu   \chi \partial_\mu F^{\mu\nu} +
  \mu_\chi  \bar{\chi} i\sigma_{\mu\nu}  \chi F^{\mu\nu},
\label{eq:gammafdm}
\eea
where $b_\chi$ and the magnetic dipole moment $\mu_\chi$ are defined as
\bea
	b_{\chi}  &=&  -\frac{\lambda_\phi^2 e}{64\pi^2 m_\phi^2} \left( 1 + \frac{2}{3}\log\frac{m_\ell^2}{m_\phi^2} \right) ,  \\
	\mu_\chi  &=&  -\frac{\lambda_\phi^2 e m_\chi}{64\pi^2 m_\phi^2}.
\eea
Note that this agrees with Ref.~\cite{Agrawal:2011ze}. The Feynman diagrams for these couplings are shown in figure~\ref{fig:vvvtri2}. The relativistic effective Lagrangian describing the interaction of the DM with quarks is
\begin{align}
	\mathcal{L}_{\rm eff} =  c^q_h \overline{\chi} \chi \overline{q}q  + c^q_{\gamma} \overline{\chi} \gamma^\mu \chi \overline{q} \gamma_\mu q
	+ c^q_{\rm md} \overline{\chi} i\sigma^{\alpha\mu} \frac{k_\alpha}{k^2} \chi \overline{q} \gamma_\mu q,
\end{align}
where 
\bea
	c^q_h        = \frac{\lambda_{\chi h} m_q}{v m_h^2}, \quad c^q_{\gamma} = e Q_q b_{\chi}  , \quad
	c^q_{\rm md} = e Q_q \mu_\chi.
\eea

We next convert the quark-level operators to nucleon-level operators and take the nonrelativistic limit. Details of the matching of operator coefficients between the quark and nucleon level operators can be found in Appendix~\ref{app:MDDM}. We thus arrive at the effective Lagrangian at the nucleon level ($N = p, n$)
\bea
	\mathcal{L}_{\rm eff} &=&  c^N_h \overline{\chi} \chi \overline{N}N +
	c^N_{\gamma} \overline{\chi} \gamma^\mu \chi \overline{N} \gamma_\mu N \nn \\
	&& + c^N_{Q} \overline{\chi}  i\sigma^{\alpha\mu}\frac{k_\alpha}{k^2} \chi \overline{N} K_\mu N \nn\\
	&& + c^N_{\mu} \overline{\chi} i\sigma^{\alpha\mu}\frac{k_\alpha}{k^2} \chi \overline{N} i\sigma^{\beta\mu} k_\beta N.
\eea
Here $K_{\mu}$ is the sum of the incoming and outgoing nucleon momenta, and the coefficients $c^N$ are related to the $c^q$ as
\begin{align}
	c^N_h &= \sum_{q = u, d, s} c^q_h \frac{m_N}{m_q} f_{Tq}^{(N)} + \frac{2}{27} f_{TG}^{(N)}   \sum_{q = c, b, t} c^q_h \frac{m_N}{m_q}, \\
	c^N_\gamma &= e  b_\chi \sum_q Q_q,		
\end{align}
and the charge and magnetic coefficients of the magnetic dipole moment are 
\bea
	c^N_{Q} = e Q_N  \mu_\chi/2m_N, \quad c^N_{\mu} = - e \tilde{\mu}_N  \mu_\chi/2m_N,
\eea
where $\tilde{\mu}_N$ is the nucleon magnetic moment, with $\tilde{\mu}_p = 2.8$ and $\tilde{\mu}_n = -1.9$.

\begin{figure}[htp]
\begin{center}
\includegraphics[width=0.35\textwidth]{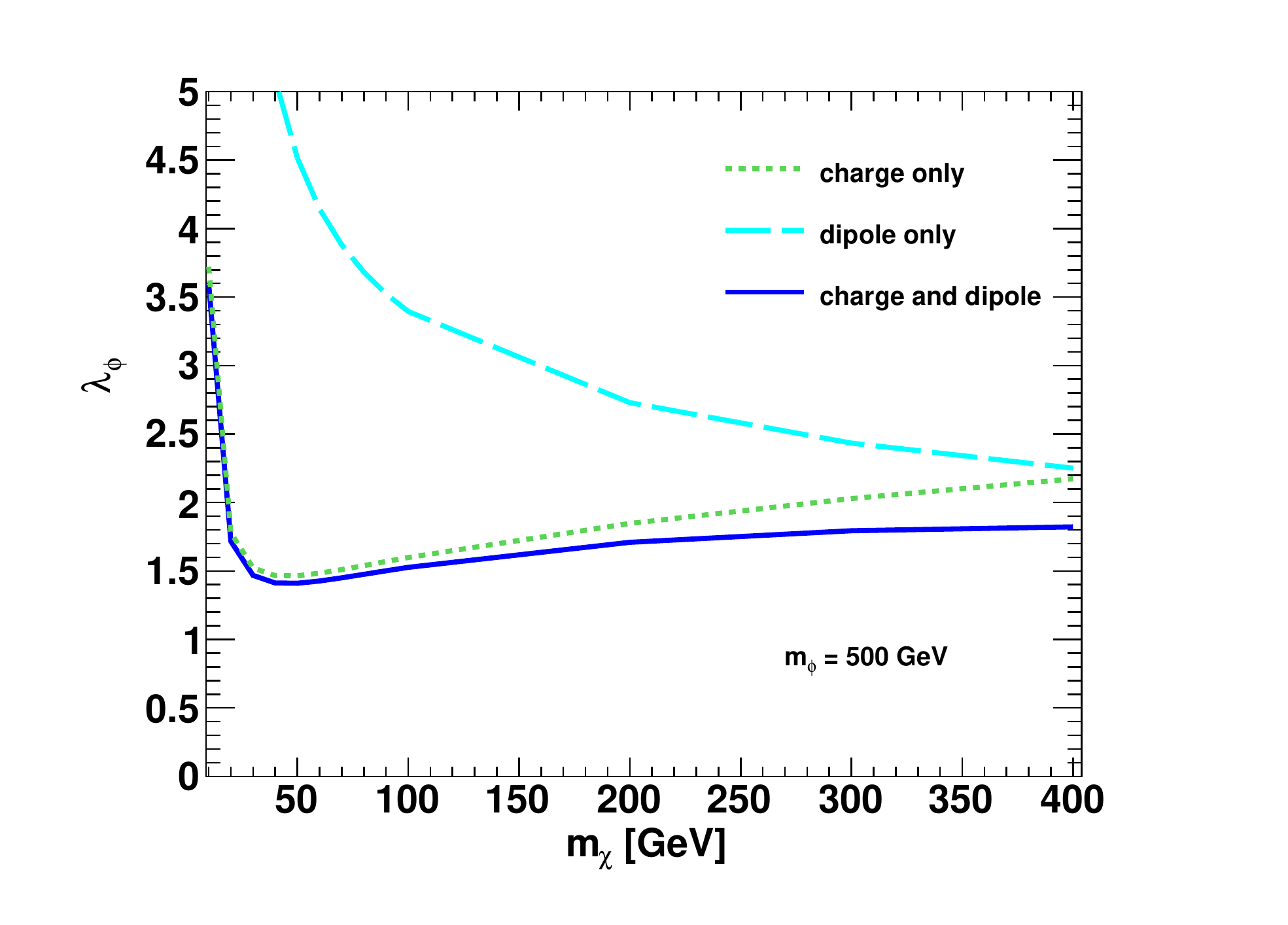} 
\caption{\small The LUX bound on the coupling $\lambda_{\phi}$ for $m_{\phi}=500$~GeV calculated using the charge term alone, the dipole term alone, and the full combination.}
\label{fig:dddipole}
\end{center}
\end{figure}

So far we have kept the magnetic dipole terms. Their momentum dependence makes it impossible to write the differential event rate as the product of the elastic cross section and the velocity integration. We calculate the differential rate numerically, and work out the exclusion limits from LUX~\cite{Akerib:2013tjd} in the presence of the dipole terms in Appendix~\ref{app:MDDM}. The result is shown in Fig.~\ref{fig:dddipole}. We find that the effect of the magnetic dipole operator is negligible compared to the charge operator in setting limits for the coupling $\lambda_{\phi}$. Based on this, for the rest of the paper we will drop the magnetic dipole contributions.

The leading (spin-independent) contribution to the nucleon matrix elements are
\begin{align}
	\langle \chi ,\, N \left| \overline{\chi} \gamma^\mu \chi \overline{N} \gamma_\mu N \right| \chi ,\, N \rangle &= 4 m_\chi m_N ,\nonumber\\
	\langle \chi ,\, N \left| \overline{\chi} \chi \overline{N}N \right| \chi ,\, N \rangle &= 4 m_\chi m_N .
	\label{eq:NmatrixF}
\end{align}
As in the scalar DM case, we define the dark matter-nucleon effective couplings
\begin{align}
	{\mathcal C}^N &=  4 m_\chi m_N c^N_{\gamma} 
	+ 4 m_\chi m_N c^N_h,
	\label{eq:NeffF}
\end{align}
where the coefficients are
\begin{align}
	c^N_{\gamma} &= - Q_N \frac{\lambda_\phi^2 e^2}{64\pi^2 m_\phi^2} \left( 1 + \frac{2}{3}\log\frac{m_\ell^2}{m_\phi^2} \right),\\
	c^N_h &=  \frac{\lambda_{\chi h} m_N}{v m_h^2} \left(\frac{2}{9} + \frac{7}{9} \sum_{q = u,\,d,\,s} f_{Tq}^{(N)}\right).
\end{align}
The total scattering cross section is then given by
\begin{equation}
	\sigma_T  = \frac{1}{16 \pi} \left( \frac{1}{m_\chi + m_p} \right)^2 \left[ Z {\mathcal C}^p + (A-Z) {\mathcal C}^n \right]^2.
\label{eq:fdd}
\end{equation}


\section{Results}
\label{sec:results}

In this section, we use the cross section formulas derived in the Sec.~\ref{sec:DD} in order to calculate the bounds on lepton-flavored DM and directly compare the regions of parameter space that have been excluded for the symmetric and asymmetric cases. Note that the full parameter space of our model is four-dimensional (with the two masses $m_{\chi}$, $m_{\phi}$ and the two couplings $\lambda_{\phi}$ and $\lambda_{\chi h}$) and therefore it is not possible to visually represent the phenomenological aspects of a full parameter scan. Instead, we choose to present the highlights in two pairs of complementary plots (for the scalar DM and fermion DM cases each), one pair where the masses are fixed at representative values and the couplings are varied, and one pair where the masses are varied, and a particular value of the couplings is chosen for each mass point. Combining the information in these plots, the reader should be able to develop an intuitive understanding for the prospects of the model in the full parameter space.

\begin{figure}[htp]
\begin{center}
\includegraphics[width=0.22\textwidth]{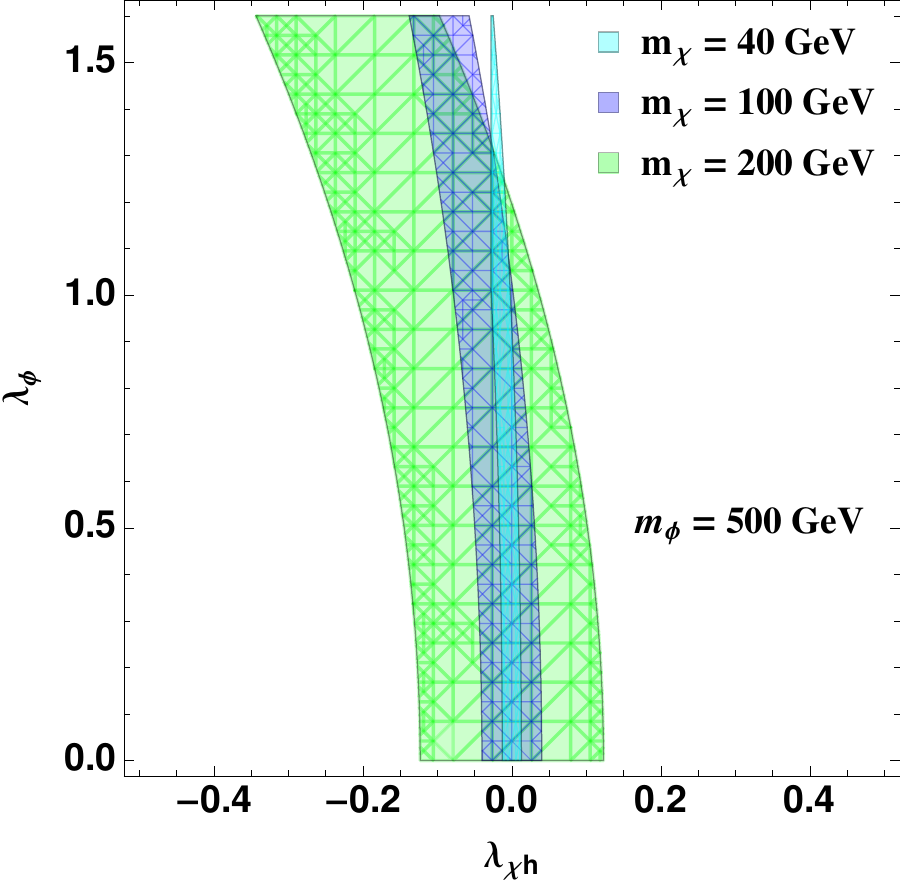} 
\includegraphics[width=0.22\textwidth]{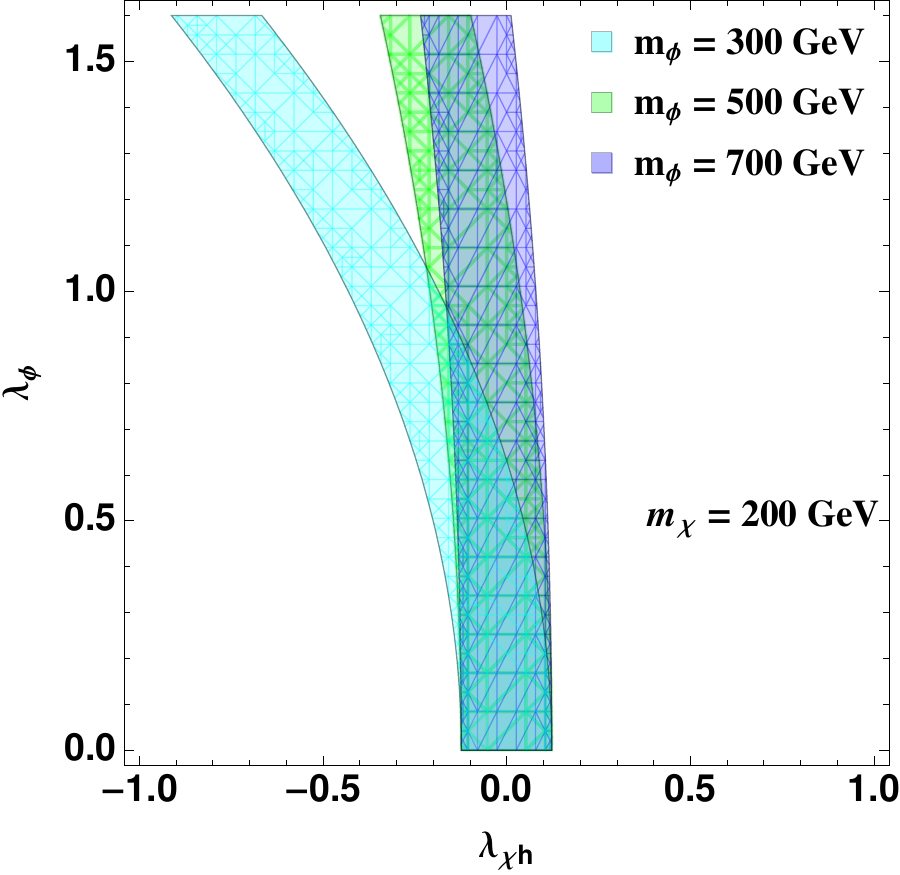}
\caption{\small The region in the $(\lambda_{\chi h}, \lambda_{\phi})$ plane for the asymmetric scalar DM case consistent with the LUX bound. (Left) $m_\phi$ fixed at 500~GeV while $m_{\chi}$ is varied. (Right) $m_\chi$ is fixed at 200~GeV while $m_{\phi}$ is varied. For $m_{\chi}=40$~GeV, the allowed region is limited to small values of $\lambda_{\chi h}$ because of the invisible Higgs decay bound.
}
\label{fig:ddsi}
\end{center}
\end{figure}

\begin{figure}[htp]
\begin{center}
\includegraphics[width=0.22\textwidth]{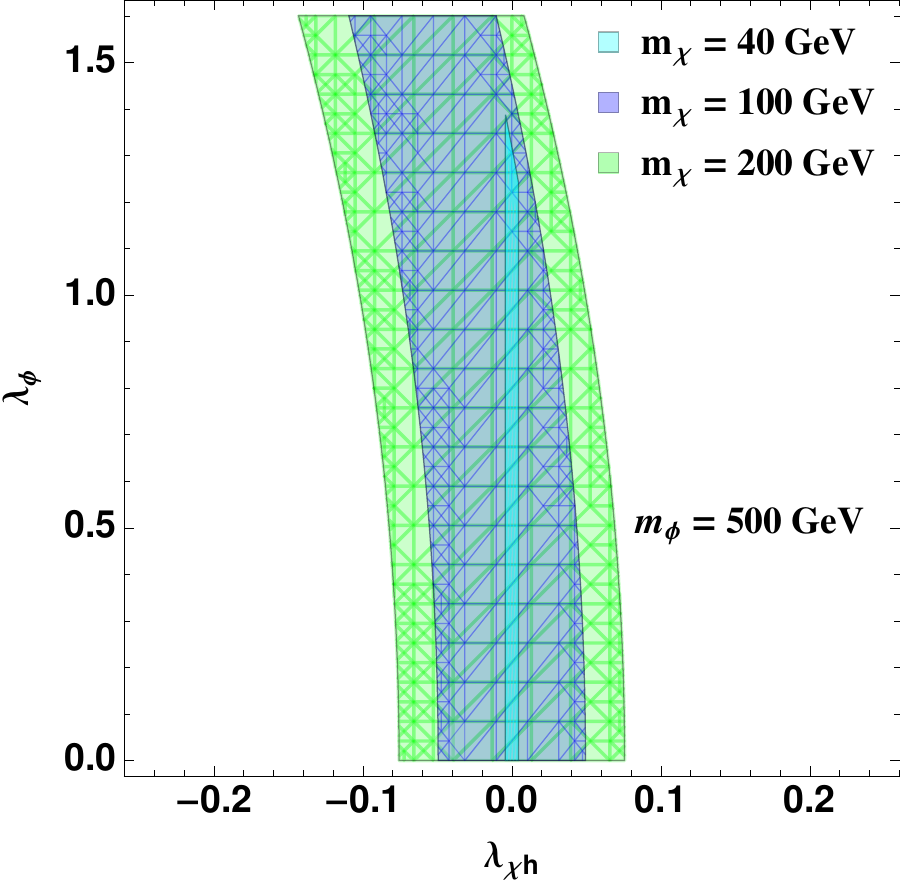} 
\includegraphics[width=0.22\textwidth]{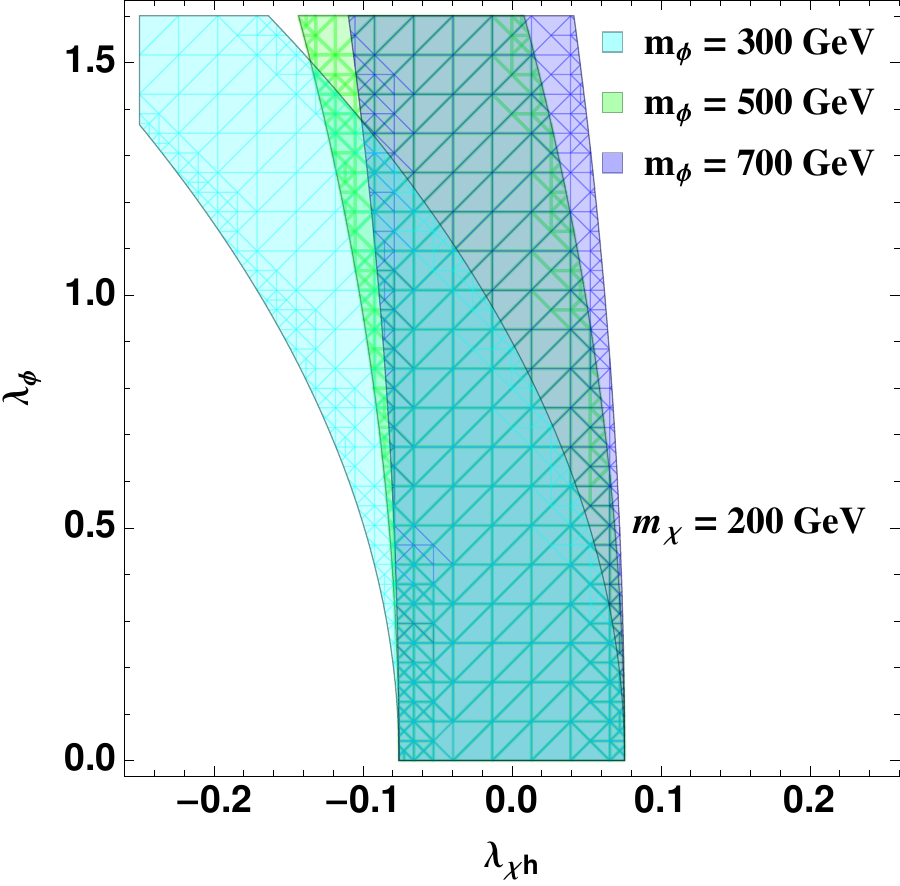} 
\caption{\small The region in the $(\lambda_{\chi h}, \lambda_{\phi})$ plane for the asymmetric fermion DM case consistent with the LUX bound. (Left) $m_\phi$ fixed at 500~GeV while $m_{\chi}$ is varied. (Right) $m_\chi$ is fixed at 200~GeV while $m_{\phi}$ is varied. For $m_{\chi}=40$~GeV, the allowed region is limited to small values of $\lambda_{\chi h}$ because of the invisible Higgs decay bound.
}
\label{fig:ddsi_fdm}
\end{center}
\end{figure}

For the asymmetric scalar and fermion DM cases, we show in Fig.~\ref{fig:ddsi}  and in Fig.~\ref{fig:ddsi_fdm} (respectively)  the regions in the $(\lambda_{\chi h}, \lambda_{\phi})$ plane for some representative choices of $m_{\chi}$ and $m_{\phi}$ that are consistent with the bounds from LUX~\cite{Akerib:2013tjd}. We also check the bounds from CREEST, CDMS-Si, and SuperCDMS~\cite{Agnese:2013rvf, Agnese:2013jaa, Angloher:2014myn}, but we find that the LUX bound dominates as long as $m_{\chi}\gsim5$~GeV. Such low values of $m_{\chi}$ are not very interesting however, as $\lambda_{\chi h}$ has to be very small in order to be consistent with the invisible Higgs decay bounds~\cite{Chatrchyan:2014tja, Aad:2014iia}, namely ${\rm BR}_{h\to\bar{\chi}\chi}<0.58$. We only plot $\lambda_{\phi}>0$ since the cross section depends only on $\lambda_{\phi}^2$, whereas the sign of $\lambda_{\chi h}$ is physical. We restrict ourselves to $|\lambda_{\chi h}|<0.25$ in the fermion DM case, since the $\chi$-Higgs coupling in this case  arises from a higher-dimensional operator which is generated at $\Lambda\gsim$~TeV [see Eq.~(\ref{eq:dim5})]. Note that the allowed parameter regions lie in a band around a curve of maximal interference. The curve of maximal interference is a parabola since the Higgs exchange amplitude scales as $\lambda_{\chi h}$ while the photon exchange amplitude scales as $\lambda_{\phi}^{2}$. In fact, the effective DM-photon coupling scales as $\lambda_{\phi}^{2}/m_{\phi}^{2}$, which explains why in the right plots the parabola moves toward the vertical axis with increasing $m_{\phi}$. While many features are similar for the scalar and fermion DM cases, one difference stands out: as can be seen the left plots, for scalar DM both the shape of the curve of maximal interference as well as the size of the allowed region around this curve depend sensitively on $m_\chi$ while for fermion DM the allowed region is much less sensitive to $m_{\chi}$. This is due to the difference between Eqs~(\ref{eq:NmatrixS}) and (\ref{eq:NmatrixF}), where in the scalar DM case the scaling of the Higgs-exchange and photon-exchange nuclear matrix elements with $m_{\chi}$ is different, while the scaling is the same in the fermion DM case.  

\begin{figure}[htp]
\begin{center}
\includegraphics[width=0.22\textwidth]{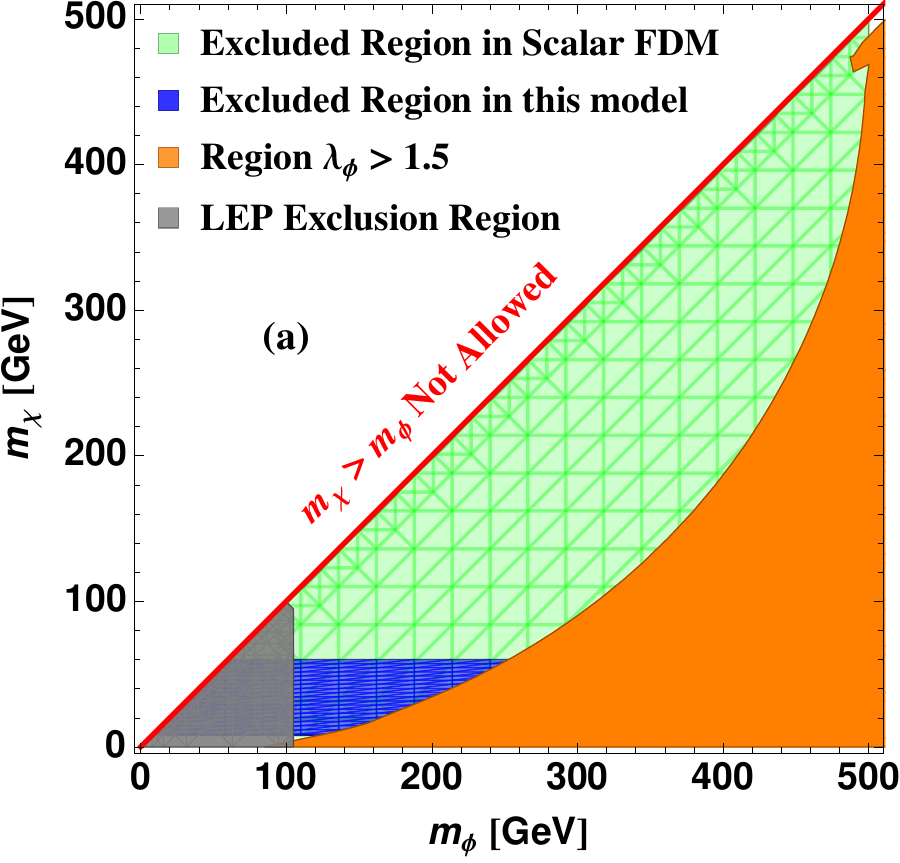} 
\includegraphics[width=0.22\textwidth]{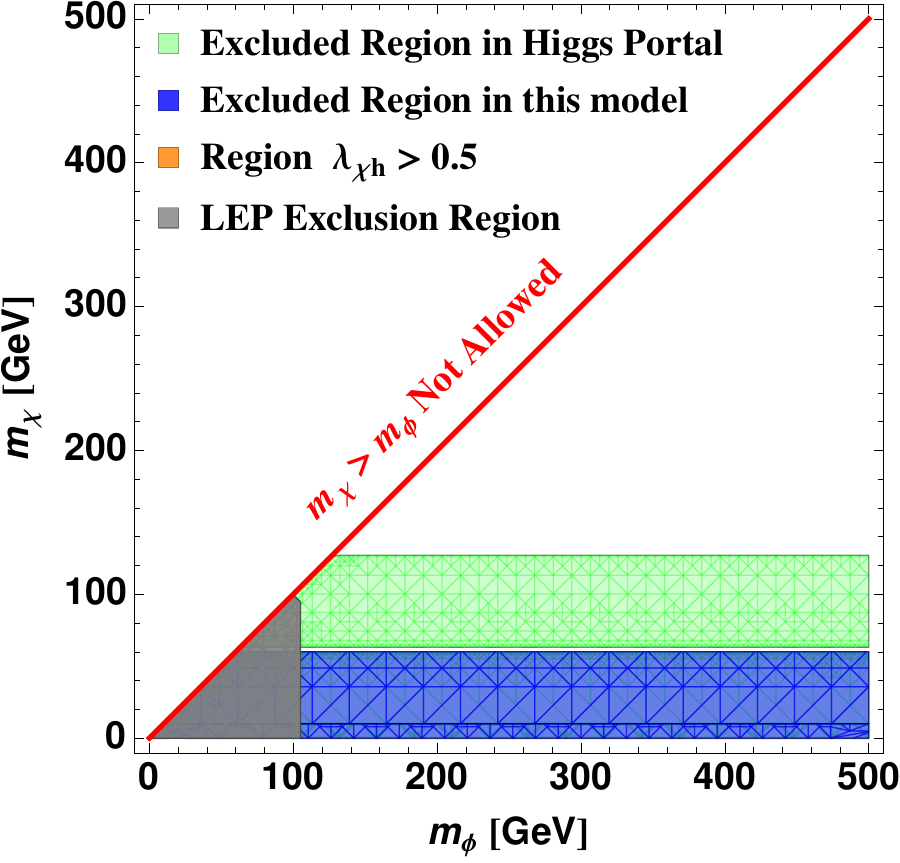}   
\caption{\small The excluded region in the $m_{\chi}$-$m_{\phi}$ plane for scalar DM. For the left plot, $\lambda_{\phi}$ is calculated point by point to give the correct relic abundance for symmetric DM. The orange region includes points where this calculated value exceeds 1.5. The green region then shows the points excluded by direct detection for symmetric DM using this value of $\lambda_{\phi}$. The blue region shows points where direct detection also excludes asymmetric DM for the same value of $\lambda_{\phi}$, and for any value of $\lambda_{\chi h}$ (subject to $|\lambda_{\chi h}|<1.5$; for $2m_{\chi}<m_{h}$, consistency with the invisible Higgs decay bound is also required). For the right plot, the roles of $\lambda_{\phi}$ and $\lambda_{\chi h}$ are reversed, and both signs of $\lambda_{\chi h}$ are used in plotting the blue region. See the main text for further details.}
\label{fig:ddrdsdm}
\end{center}
\end{figure}

\begin{figure}[htp]
\begin{center}
\includegraphics[width=0.22\textwidth]{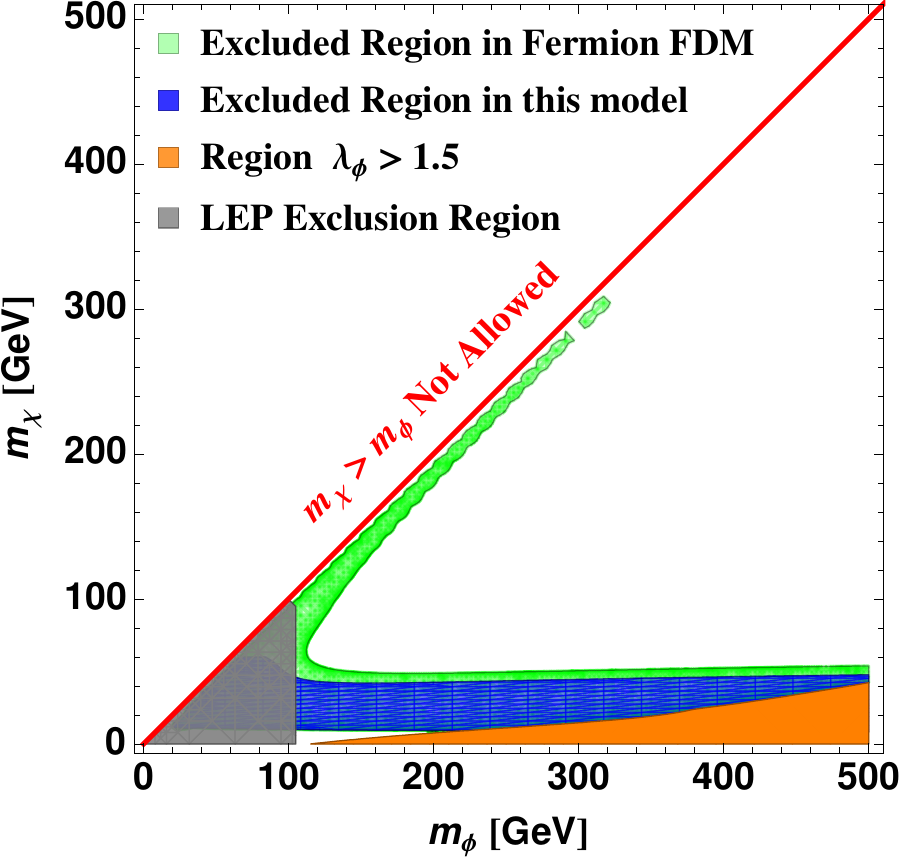} 
\includegraphics[width=0.22\textwidth]{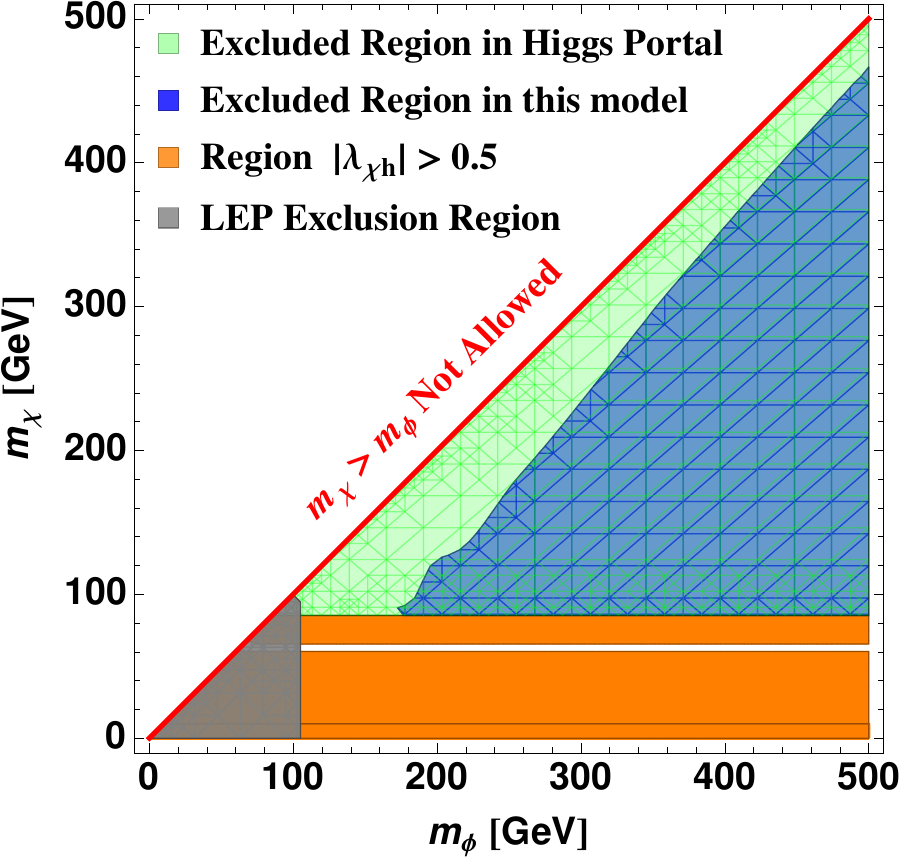}   
\caption{\small The excluded region in the $m_{\chi}$-$m_{\phi}$ plane for fermion DM. For the left plot, $\lambda_{\phi}$ is calculated point by point to give the correct relic abundance for symmetric DM. The orange region includes points where this calculated value exceeds 1.5. The green region then shows the points excluded by direct detection for symmetric DM using this value of $\lambda_{\phi}$. The blue region shows points where direct detection also excludes asymmetric DM for the same value of $\lambda_{\phi}$, and for any value of $|\lambda_{\chi h}|<0.5$ (for $2m_{\chi}<m_{h}$, consistency with the invisible Higgs decay bound is also required). For the right plot, the roles of $\lambda_{\phi}$ and $\lambda_{\chi h}$ are reversed, and both signs of $\lambda_{\chi h}$ are used in plotting the blue region. See the main text for further details.}
\label{fig:ddrdfdm}
\end{center}
\end{figure}

Next, we contrast the regions in the parameter space that can be consistent with the LUX bound for symmetric and antisymmetric lepton-flavored DM as a function of the masses $m_{\chi}$ and $m_{\phi}$. In the left plot of Figs.~\ref{fig:ddrdsdm} and \ref{fig:ddrdfdm} (for scalar and fermion DM, respectively), we start by calculating for any point in the $m_{\chi}$-$m_{\phi}$ plane the value of $\lambda_{\phi}$ that gives rise to the correct relic density in the symmetric DM case (for details of the relic abundance calculation, see Appendix~\ref{app:relic}). For the symmetric DM case, we then check whether this parameter point is excluded by direct detection, keeping $\lambda_{\chi h}=0$, since for the symmetric case the two channels add incoherently so any finite value of $\lambda_{\chi h}$ only strengthens the direct detection constraint. Next, for the same value of $\lambda_{\phi}$, we check whether there is any value of $\lambda_{\chi h}$ (within the interval [-1.5,\ 1.5] for scalar DM and [-0.5,\ 0.5] for fermion DM, and consistent with the invisible Higgs decay bound if $2m_{\chi}<m_{h}$) for which asymmetric DM can be consistent with the direct detection bound. In the second plot (right), we exchange the roles of $\lambda_{\chi h}$ and $\lambda_{\phi}$ and repeat the same procedure, in other words $\lambda_{\chi h}$ is now fixed at the value which gives the correct relic abundance for the symmetric DM (both signs are considered) at any value of $m_{\chi}$ and $m_{\phi}$ (subject to the same constraints as mentioned above), and for antisymmetric DM $\lambda_{\phi}$ is allowed to float in looking for consistency with the direct detection bound. Note that we have excluded the regions $m_{\phi}<105$~GeV in these plots due to $\phi$-pair production bounds from LEP. This is only meant as a conservative approximation to the LEP bound, however the direct search bounds from the LHC (such as stau searches) will rule out this region in any case and extend further, and for this reason the lowest $m_{\phi}$ regions should not be taken too seriously. A full analysis of the LHC constraints will be studied in upcoming work, but it is outside the scope of this paper due to the large number of LHC searches that need to be recast. Since pair production cross sections of noncolored particles (especially scalars) fall off very rapidly however, we do not expect the inclusion of LHC bounds to drastically change plots~\ref{fig:ddrdsdm} and \ref{fig:ddrdfdm}.

While choosing either $\lambda_{\phi}=0$ or $\lambda_{\chi h}=0$ for the symmetric case in Figs.~\ref{fig:ddrdsdm} and \ref{fig:ddrdfdm} may appear to be somewhat arbitrary, this in fact allows us to fully map out the exclusion region from direct detection bounds, in the following sense: If a point in the $m_{\chi}$-$m_{\phi}$ plane is excluded by LUX in both Figs.~\ref{fig:ddrdsdm} and \ref{fig:ddrdfdm} in the symmetric case, it is ruled out even when both couplings are allowed to vary, subject to the relic abundance constraint, as we will now argue. Interference is absent in the symmetric case in calculating the scattering rate for direct detection, which therefore can be written as $C_{DD,\phi}(\lambda_{\phi}^{2})^{2}+C_{DD,h}\lambda_{\chi h}^{2}$ for some constants $C_{DD,\phi}$ and $C_{DD,h}$. Similarly, the cross section of DM annihilation relevant for the relic abundance calculation can also be written as $C_{RA,\phi}(\lambda_{\phi}^{2})^{2}+C_{RA,h}\lambda_{\chi h}^{2}$, for the same reason\footnote{There is a caveat here that the reaction $\chi\chi\to\tau\tau$ does in fact have a cross term between $\phi$ and $h$ exchange. However, the Yukawa coupling of the $\tau$ is small enough that this term can be neglected for all practical purposes.}. Thus, for a given mass point, obtaining the correct relic abundance constrains the model to lie on an ellipse in the $\lambda_{\phi}^{2}$-$\lambda_{\chi h}$ plane, with the major axis pointing along either the $\lambda_{\phi}^{2}$ or the $\lambda_{\chi h}$ axis. Moreover, the contours corresponding to constant scattering rate in a direct detection experiment are also ellipses with their major axis pointed along either the $\lambda_{\phi}^{2}$ or the $\lambda_{\chi h}$ axis. Thus, as one moves around the ellipse for obtaining the correct relic abundance, one will always find the point with the smallest direct detection scattering rate where the family of ellipses from direct detection are tangent to the ellipse from relic abundance. Since both ellipses are pointed along one of the coordinate axes, this will happen on one of the coordinate axes, thus there can be no point with both couplings nonzero that can result in a smaller scattering rate for direct detection than at the points with one of the couplings equal to zero.

There are many interesting features in Figs.~\ref{fig:ddrdsdm} and \ref{fig:ddrdfdm}, which we now go over in detail. First of all, note that for symmetric DM, the entire parameter region is excluded for fermion DM with negligible FDM coupling (right plot in Fig.~\ref{fig:ddrdfdm}, apart from a very narrow band near the Higgs resonance region, where $\lambda_{\chi h}$ can be very small). This exclusion extends all the way down to zero mass due to the invisible Higgs bound. Similarly, scalar DM with negligible Higgs coupling (left plot in Fig.~\ref{fig:ddrdsdm}) is also ruled out for a DM mass above 8~GeV, below which direct detection experiments lose sensitivity. The nearly-complete exclusion for these two cases is due to relic abundance requiring a very large coupling due to suppressions in the amplitude. Scalar DM with FDM interactions annihilates to leptons, so the s-wave annihilation is chirally suppressed [see Eq.~{\ref{eq:relSDMa})] and therefore p-wave annihilation dominates. Fermion DM that annihilates through Higgs exchange is also velocity suppressed. In both cases, turning on both couplings for asymmetric DM opens up regions of parameter space that can be consistent with all constraints. In particular, for scalar DM, the only region that is ruled out is for $2m_{\chi}<m_{h}$ where the invisible Higgs decay bound forces $\lambda_{\chi h}$ to be very small such that the interference cannot be very effective. For fermion DM where we set $\lambda_{\chi h}$ by the relic abundance in the symmetric case (Fig.~\ref{fig:ddrdfdm}, right), the large $m_{\phi}$ region is ruled out even for the asymmetric case, because the effective DM-photon coupling scales as $\lambda_{\phi}^{2}/m_{\phi}^{2}$, so a value of $\lambda_{\phi}$ of order one is not strong enough to cancel the very large Higgs exchange contribution in direct detection.

There are also a few interesting features in the left plot of Fig.~\ref{fig:ddrdfdm}. For symmetric DM, the exclusion region extends both to large $m_{\phi}$ for light $m_{\chi}$, as well as to relatively large $m_{\chi}$ when $m_{\phi}-m_{\chi}$ is small. The former region is ruled out because both direct detection and relic abundance depend on $\lambda_{\phi}$ and $m_{\phi}$ in the same way, thus the direct detection constraint does not weaken even at large $m_{\phi}$. The latter region is ruled out because the loop that gives rise to the effective DM-photon coupling is enhanced in this kinematic regime, and therefore the direct detection bound is stronger than one would naively expect. In a way similar to Fig.~\ref{fig:ddrdsdm} (left), the excluded region for asymmetric DM is basically due to the invisible Higgs decay bound, which forces $\lambda_{\chi h}$ to remain small, and therefore makes the interference ineffective. Also similar to Fig.~\ref{fig:ddrdsdm} (left), the region $m_{\chi}<10~$GeV is not excluded because direct detection experiments lose sensitivity at such low recoils.


\section{Conclusion and Outlook}
\label{sec:conclusions}

We have introduced the scenario of lepton-flavored asymmetric dark matter, where the same mechanism that generates a lepton asymmetry at high scales also generates a DM asymmetry, and we have studied the prospects of this scenario for direct detection experiments. In particular, we have emphasized the fact that the interactions present in the model lead to both Higgs and photon exchange in direct detection, and that the corresponding amplitudes are naturally of the right size such that interference can be important, leading to a significant weakening in the bounds reported by direct detection experiments. We have contrasted the regions of parameter space excluded by the null results of direct detection experiments for this scenario with the parameter space of the same model where no DM asymmetry is generated, and where therefore the interference effects cancel out once the scattering of both the DM particle and its antiparticle off of nuclei are taken into account. In particular, we showed that in the symmetric case with fermion DM where the Higgs exchange dominates, the parameter space is entirely ruled out except for a narrow Higgs resonance window, while in the asymmetric case a large fraction of the parameter space is still allowed. The same conclusion also holds for scalar DM with a mass above 8~GeV when the FDM interaction dominates.

Turning to prospects of this model for future experiments, we note that the presence of interference in direct detection can be confirmed by separately determining the DM scattering rate off of protons and neutrons. This can be achieved in the next generation of direct detection experiments if more than one experiment with a nonidentical active detector material can observe a signal, since the ratios of protons to neutrons in the nuclei of the active materials will then be different. A separate measurement of the scattering rates from protons and neutrons can then be used to solve for $\lambda_{\phi}$ and $\lambda_{\chi h}$.

While indirect detection signals are absent for asymmetric DM, the collider phenomenology of our model is identical to the symmetric case. The discovery prospects in the multilepton final state at the LHC were studied in Ref.~\cite{Agrawal:2011ze} at which point no collider constraints were available. It would now be interesting to study the constraints imposed on the lepton-flavored dark matter model by translating the searches performed by ATLAS and CMS in the dilepton   and multilepton   final states with and without transverse missing energy. Due to the multiplicity of such analyses this was outside the scope of this paper, but these constraints will be studied in upcoming work.

In this paper we considered it sufficient to simply outline the details of a model which would lead to the generation of a DM asymmetry during high-scale leptogenesis, and to remark that an order one coupling for the FDM interaction would then efficiently annihilate the symmetric part of the DM particles. In future work we plan to take up this question in greater quantitative detail and calculate the energy density left over in the asymmetric DM as a function of the parameters of the UV model.

\section*{Acknowledgements}

We would like to thank Prateek Agrawal, Zackaria Chacko, Jay Hubisz, Gordan Krnjaic, Flip Tanedo, Wei Xue and Kathryn Zurek for helpful discussion and valuable comments. C.K. would also like to thank the Aspen Center for Physics (supported by the National Science Foundation under Grant No. PHYS-1066293) as well as the Perimeter Institute for Theoretical Physics (supported by the Government of Canada through Industry Canada and by the Province of Ontario through the Ministry of Research and Innovation), where part of this work was completed, for their hospitality. The research of the authors is supported by the National Science Foundation under Grants No. PHY-1315983 and No. PHY-1316033.


\appendix


\begin{widetext}
	
	
\section{Relic Density Calculations}
\label{app:relic}

Since we wish to compare the parameter space consistent with direct detection bounds when the DM is asymmetric with the usual thermal relic case, we need to calculate the relic abundance in our model when no asymmetry is generated at high scales. Here we list the results of this calculation, for the scalar and fermion DM cases.

\subsection{Scalar dark matter}

The coupling to the Higgs gives rise to the following annihilation channels, with their respective cross sections:
\bit
\item $\chi\chi \to f \bar{f}$:
\bea
	&&\sigma v_{\rm{rel}}  = \frac{1}{4\pi} N_C^f \lambda_{\chi h}^2  \beta_f^3 m_f^2 
	\frac{1}{(s - m_h^2)^2 + m_h^2 \Gamma_h^2},
\eea
where  $\beta_i = \sqrt{1 - 4m_i^2/s}$.
\item $\chi\chi \to VV$:
\bea
	&&\sigma v_{\rm{rel}} = \frac{1}{1+\delta_{VZ}}\frac{1}{8\pi} \lambda_{\chi h}^2
	\beta_V
	\frac{s}{(s - m_h^2)^2 + m_h^2 \Gamma_h^2}\left(1 - 4\frac{ m_V^2}{s} + 12 \frac{ m_V^4}{s^2}\right),
\eea
where $\delta_{VZ}$ is 1 for the $Z$ boson and 0 for the $W$ boson. 
\item $\chi\chi \to hh$:
\bea
\sigma v_{\rm rel}
&=& \frac{\lambda_{\chi h}^2 \beta_h}{	16\pi s}
    \left[ \; \left(  \frac{ s+ 2 m_h^2}{s-m_h^2}  \right)^2     
        + \frac{2 \lambda_{\chi h}^2v^4}{(m_\chi^2 - t_-)(m_\chi^2 - t_+)} 
	+ 4 \lambda_{\chi h} \frac{v^2 }{s\beta_\chi \beta_h} \left(  \frac{ s+ 2 m_h^2}{s-m_h^2} - \frac{\lambda_{\chi h}v^2}{s-2m_h^2} \right)
          \log \left| \frac{m_\chi^2 - t_+}{m_\chi^2 - t_-} \right|  
    \right] \,,
\label{svhh}
\eea
where $t_\pm =  m_\chi^2 + m_h^2 - \frac{1}{2}s (1  \mp  \beta_\chi \beta_h)$.

\eit

The FDM coupling of Eq.~(\ref{eq:FDMvertexS}) gives rise to the annihilation channel ${\chi}^*\chi\to\ell^{+}\ell^{-}$. The cross section can be written as
\bea
	&&\sigma v_{\rm{rel}} = a + b v^2,
\eea
where 
\bea
	a &=& \frac{\lambda_{\phi}^4}{16 \pi}\frac{ m_f^2 }{(m_\phi^2 + m_\chi^2 - m_f^2)^2} \left( 1 - \frac{m_f^2}{m_\chi^2}\right)^\frac32,\label{eq:relSDMa}\\
	b &=& \frac{\lambda_{\phi}^4}{48 \pi} \frac{m_\chi^2}{(m_\phi^2 + m_\chi^2)^2}.
\eea
Note that the $s$-wave contribution is chirality-suppressed.

\subsection{Fermion dark matter}

The coupling to the Higgs gives rise to the following annihilation channels, with their respective cross sections:
\bit
\item $\chi\chi \to f \bar{f}$:
\bea
	&&\sigma v_{\rm{rel}}  = \frac{1}{8\pi} N_C^f  \frac{\lambda_{\chi h}^2 m_f^2}{v^2} \beta_\chi^2 \beta_f^3
	\frac{s}{(s - m_h^2)^2 + m_h^2 \Gamma_h^2}.
\eea

\item $\chi\chi \to VV$:
\bea
	&&\sigma v_{\rm{rel}} = \frac{1}{1+\delta_{VZ}}\frac{1}{16\pi}  \frac{\lambda_{\chi h}^2}{v^2}  \beta_\chi^2 \beta_V
	\frac{s^2}{(s - m_h^2)^2 + m_h^2 \Gamma_h^2}\left(1 - 4\frac{ m_V^2}{s} + 12 \frac{ m_V^4}{s^2}\right).
\eea
\item $\chi\chi \to hh$:
\bea
\sigma v_{\rm rel}
&=& 
	\frac{\lambda_{\chi h} ^2 \beta_h \beta_{\chi}^2}{32 \pi  v^2}
	\left[
	\left(\frac{s+2 m_{h}^2}{s-m_{h}^2}\right)^2
	+
	\frac{8 \lambda_{\chi h}  m_{\chi} v \left(s+2 m_{h}^2\right)}{\beta_{\chi}^2 s \left(s-m_{h}^2\right)}
	-
	\frac{2 \lambda_{\chi h} ^2 v^2 \left(3 m_{h}^4-16 m_{h}^2 m_{\chi}^2+2 m_{\chi}^2 \left(s+8 m_{\chi}^2\right)\right)}{\beta_{\chi}^2 s \left(m_{h}^4-4 m_{h}^2 m_{\chi}^2+m_{\chi}^2 s\right)}
	\right. \nn\\
	&&\left. +
	\frac{2 v \lambda_{\chi h} }{\beta_h \beta_{\chi}^3 s^2}
	\left(
	\frac{v \lambda_{\chi h} \left(6 m_{h}^4-4 m_{h}^2 \left(s+4 m_{\chi}^2\right)-32 m_{\chi}^4+16 m_{\chi}^2 s+s^2\right)}{s-2 m_{h}^2}
	-
	\frac{2  m_{\chi} \left(s+2 m_{h}^2\right) \left(2 m_{h}^2-8 m_{\chi}^2+s\right)}{m_{h}^2-s}
	\right)
	\right. \nn\\
	&&\left.
	\log  \left| \frac{m_\chi^2 - t_+}{m_\chi^2 - t_-} \right|  
	\right].
\label{svhhf}
\eea

\eit

The FDM coupling of Eq.~(\ref{eq:FDMvertexF}) gives rise to the annihilation channel $\bar{\chi}\chi\to\ell^{+}\ell^{-}$. Unlike the scalar DM case, here the cross section is dominated by the $s$-wave:
\bea
	&&\sigma v_{\rm{rel}} = \frac{\lambda_{\phi}^4}{32 \pi}\frac{ m_\chi^2 }{(m_\chi^2 + m_\phi^2 - m_f^2)^2} \sqrt{ 1 - \frac{m_f^2}{m_\chi^2}}.
\eea


\section{The effective DM-photon coupling}
\label{app:loop}

\subsection{Scalar dark matter}

The DM-photon interaction induced at one-loop has the form 
\bea
		{\mathcal L}_{\rm eff} = i b_\chi \partial_\mu \chi^* \partial_\nu \chi F^{\mu\nu}
\eea
where
\bea
 b_\chi &=& - \frac{e \lambda^2}{16\pi^2}  \int_0^1 dy\bigg[ \frac{y^3( \Delta_0(6-4y)+y(m_\chi^2(1-y)^2+m_\phi^2))}{3 \Delta_0^2} 
 - (m_\phi^2 \leftrightarrow m_\ell^2)\bigg], \label{complexscalarbx}
 \eea
and 
 \bea
    \Delta_0 \equiv  m_\phi^2 y + (1 - y) m_\ell^2 - y(1-y) m_\chi^2.
 \eea
In the limit $m_{\chi}\ll m_{\phi}$ and $m_{\ell}\ll m_{\phi}$, $b_\chi$ is given to leading order by
\bea
	b_\chi = \frac{\lambda^2 e}{16 \pi^2 m_{\phi}^{2}} \left(1-\frac{4}{3}\log \left( \frac{m_\ell}{m_\phi} \right)\right).
\eea

\subsection{Fermion dark matter}

The DM-photon interaction induced at one-loop has the form 
\bea
  {\mathcal L}_{\rm eff} &=& b_\chi  \bar{\chi} \gamma_\nu   \chi \partial_\mu F^{\mu\nu} +
  \mu_\chi  \bar{\chi} i\sigma_{\mu\nu}  \chi F^{\mu\nu},
\label{eq:gammafdm2}
\eea
where
\bea
    \mu_\chi &=& -\frac{ie \lambda^2}{64\pi^2 } \int^1_0 d y\  2 m_\chi  \frac{ y(1-y)}{\Delta_0} ,\\
    b_{\chi} & = & 		-\frac{ie \lambda^2}{64\pi^2 } \int^1_0 d y\  \frac{1}{6} y    \left[  (y^2 -3y) \left(\frac{1}{\Delta_0}-\frac{1}{\Delta_0^\prime}\right) +  ( y^2-6y+3) \frac{1}{\Delta_0} +  (y^2 -3y) \frac{m_\ell^2}{\Delta_0^2}\right],
\eea
and 
\bea
	\Delta_0
	       &=&  y m_\ell^2 + (1-y)m_\phi^2 -  y(1 - y) m_\chi^2 , \\ 
	   	\Delta^\prime_0
	   	       &=&  y m_\phi^2 + (1-y)m_\ell^2 -  y(1 - y) m_\chi^2.
\eea
In the limit $m_{\chi}\ll m_{\phi}$ and $m_{\ell}\ll m_{\phi}$, $\mu_\chi$ and $b_{\chi}$ are given to leading order by
\bea
\mu_{\chi} &=& -\frac{ e \lambda^2 m_{\chi} }{64\pi^2 m_\phi^2},\\
b_{\chi} &=& -\frac{i\lambda^2 e }{64\pi^2 m_\phi^2}\left(1 + \frac23 \log\frac{m_\ell^2}{m_\phi^2}\right).
\eea


\section{Including the magnetic dipole interaction in direct detection}
\label{app:MDDM}

In this appendix we report the calculation details related to obtaining the direct detection bound for the fermion DM case when the dipole interaction is taken into account. The one-loop induced effective Lagrangian at the nucleon level is
\bea
	\mathcal{L}_{\rm eff} &=&   
	c^N_{\gamma} \overline{\chi} \gamma^\mu \chi \overline{N} \gamma_\mu N  + c^N_{Q} \overline{\chi}  i\sigma^{\alpha\mu}\frac{k_\alpha}{k^2} \chi \overline{N} K_\mu N   + c^N_{\mu} \overline{\chi} i\sigma^{\alpha\mu}\frac{k_\alpha}{k^2} \chi \overline{N} i\sigma^{\beta\mu} k_\beta N,
\eea
where
\bea
	c^N_\gamma &= e Q_N b_\chi,	 \quad c^N_{Q} = e Q_N  \mu_\chi, \quad c^N_{\mu} = - e \tilde{\mu}_N  \mu_\chi.
\eea
Due to the nontrivial momentum dependence of these operators, we cannot directly use the elastic cross section bounds reported by the direct detection experiments. We thus proceed to calculate the differential rate and the event rate based on the parameters of the direct detection experiment and on the local DM velocity distribution. The differential scattering rate is given by
\bea
	\frac{\rm{d} {\mathcal R} }{\rm{d} E_R} 
	= 
	N_T\frac{\rho_\chi}{m_\chi}
	\int_{\vmin} \, v f(\bv)\, \frac{\rm{d} \sigma}{\rm{d} E_R} d^3 \bv,
\eea
where $f(\bv)$ is the local dark matter velocity distribution, $\rho_{\chi}$ is the local DM density  (taken to be $0.3$ GeV/cm$^3$), and $N_T $ denotes the number of target nuclei per unit mass of the detector.

Let us start with the differential scattering cross section $\frac{\rm{d} \sigma}{\rm{d} E_R}$. In the nonrelativistic limit, the leading contributions~\cite{Fitzpatrick:2012ix, Gresham:2014vja} to the relativistic nucleon-level operators are
\begin{align}
	\langle \chi ,\, N \left| \overline{\chi} \gamma^\mu \chi \overline{N} \gamma_\mu N \right| \chi ,\, N \rangle &= 4 m_\chi m_N ,\\
	\langle \chi ,\, N \left| \overline{\chi}  i\sigma^{\alpha\mu}\frac{k_\alpha}{k^2} \chi \overline{N} K_\mu N \right| \chi ,\, N \rangle &= 
	 4 m_N^2  + 16i  {m_N^3 m_\chi \over k^2 } \vec{v}^\perp \cdot \left(  {\vec{k} \over m_N } \times \vec{S}_\chi \right) ,\\
	\langle \chi ,\, N \left|  \overline{\chi} i\sigma^{\alpha\mu}\frac{k_\alpha}{k^2} \chi \overline{N} i\sigma^{\beta\mu} k_\beta N \right| \chi ,\, N \rangle &= 16 m_\chi m_N \frac{m_N^2}{k^2} \left(  {\vec{k} \over m_N} \times \vec{S}_\chi \right) \cdot \left(  {\vec{k} \over m_N} \times \vec{S}_N \right).
\end{align}
At the nuclear level, taking the nuclear responses into account, and averaging over spins, we get
\bea
	\frac{1}{2(2 J + 1)} &&\frac1{(4 m_\chi m_T)^2}\sum_{\rm spin}|{\mathcal M}|^2_{\rm nuclear} =e^2 b_\chi^2  \W_M^{(p,p)} +  e^2  \mu_\chi^2 \left( \frac{\vec{v}^{2}}{\vec{k}^{\,2}} - {1 \over 4 \mu_{\chi T}^2} + {1 \over 4 m_\chi^2} \right) \W_M^{(p,p)} \\
	&+&    {e^2 \mu_\chi^2 \over m_N^2} \left[\W_\Delta^{(p,p)}- \mag_n \W_{\Delta \Sigma'}^{(p,n)}-\mag_p \W_{\Delta\Sigma'}^{(p,p)} + {1 \over 4} \left( \mag_p^2 \W_{\Sigma'}^{(p,p)}+2 \mag_n \mag_p \W_{\Sigma'}^{(p,n)}+\mag_n^2 \W_{\Sigma'}^{(n,n)}\right) \right],
	\label{eq:dipolenuclear}
\eea
where $\W_i^{(N,N)}$ are nuclear response functions with nuclear spin average factor $\frac{1}{2 J + 1}$ included, defined in Refs.~\cite{Fitzpatrick:2012ix, Gresham:2014vja}. We use the shell model to write the magnetic moment of a nucleus as
\beq
\mag_T = 2 \mag_p \langle S_p \rangle+2 \mag_n \langle S_n \rangle+ \langle L_p \rangle. \label{eq: mag moment}
\eeq
In the $q^2 \rightarrow 0$ limit, the term in square brackets in Eq.~(\ref{eq:dipolenuclear}) goes to ${J+1 \over 6 J} \mag_T^2$, while $\W_M^{(p,p)}$ becomes $Z^2$. Eq.~(\ref{eq:dipolenuclear}) thus simplifies to
\bea
	\frac{1}{2(2 J + 1)} &&\frac1{(4 m_\chi m_T)^2}\sum_{\rm spin}|{\mathcal M}|^2_{\rm nuclear} = 
	e^2 \mu_\chi^2 \left[{\vec{v}^{\,2} \over \vec{k}^{\,2}} -{ 1\over 4} \left(\frac{2}{m_T m_\chi}
	+\frac{1}{m_T^2}\right)\right] Z^2 F^2(A; k^2) +  e^2 \mu_\chi^2\frac{J+1}{3 J} \frac{\tilde{\mu}_T^2}{2 m_N^2},
\eea
where $F(A; k^2)$ is the Helm form factor. 
With this, the differential cross section becomes (consistent with Refs.~\cite{Barger:2010gv, Gresham:2013mua})
\bea
	\frac{\rm{d} \sigma}{\rm{d} E_R} = \frac{m_T}{2\pi v^2} \left\{  e^2 b_\chi^2 Z^2 F^2(A; k^2) + 
	e^2 \mu_\chi^2\left[{\vec{v}^{\,2} \over \vec{k}^{\,2}} -{ 1\over 4} \left(\frac{2}{m_T m_\chi}
		+\frac{1}{m_T^2}\right)\right] Z^2 F^2(A; k^2) +  e^2 \mu_\chi^2\frac{J+1}{3 J} \frac{\tilde{\mu}_T^2}{2 m_N^2}
		\right\}.
\eea

We next turn our attention to modeling detector effects of the direct detection experiment. In particular, we have to take into account that the measured energy is only part of the true recoil energy $E_R$, that the experiment has a finite energy resolution and that the analysis involves cuts, the efficiencies of which will enter the calculation of the differential rate.

The LUX experiment uses the direct scintillation (S1) and ionization signals (S2) to reject backgrounds. Both the S1 and S2 signals are detected by arrays of photomultipliers (PMTs), and measured in numbers of photoelectrons (PE). The expected number of photoelectrons~\cite{Aprile:2011hi,Akerib:2013tjd} is
\bea
	\nu(E_R) = E_R \times {\mathcal L}_{eff} \times L_y S_{\rm nr}/S_{\rm ee}.
\eea
In using this formula, we take the values for the scintillation efficiency ${\mathcal L}_{eff}$ and energy dependent absolute light yield $L_y S_{\rm nr}/S_{\rm ee}$ (with scintillation quenching factors for electron and nuclear recoils included) from page 25 of the slides at \url{http://luxdarkmatter.org/talks/20131030_LUX_First_Results.pdf}.

The smearing function has a mean $n$ and variance $\sqrt{n}\sigma_{\rm PMT}$ with $\sigma_{\rm PMT} = 0.37 $ PE. Since the analysis uses the lower half of the signal band, the cut efficiency is taken to be $50\%$~\cite{Akerib:2013tjd}. The number of signal events thus becomes
\bea
N = {\rm Ex}\times \int^{S1_{up}}_{S1_{\rm low}} {\rm d} S1\  \mathcal{E}(S1) \sum_{n=1}^\infty {\rm Gauss}(S1|n,\sqrt{n}\sigma_{\rm PMT})
\int^\infty_0 {\rm d} E_R\  {\rm Poisson}(n|\nu(E_R))  \frac{{\rm d}\mathcal{R}}{{\rm d}E_R} .
\eea
where the $S1$ integration range is 2 PE$\le S_1\le$30 PE, and ${\rm Ex}$  denotes the experimental exposure, taken to be $85.3 \times 118$ kg-days. 

Putting everything together, we calculate the probability for the signal plus background to have given rise to no more than one event (as was observed by LUX). This is given as
\bea
	{\mathcal L} = \sum_{k = 0}^1 \int {\rm d} \mu_B {\rm Gauss}(\mu_B|N_B,\sigma_B) {\rm Poisson}(k|N_B + N_S),
\eea
where $N_S$ is the expected number of the signal events, $N_B$ is the expected number of background events and $\sigma_B$ is its variance. We take the latter two parameters to be $0.64 \pm 0.16$. We then use ${\mathcal L}$ to set to bound on $N_S$ at $90\%$ confidence level, which can then be translated to a bound in terms of the model parameters.


\end{widetext}


\bibliography{paper}

\end{document}